\documentclass[%
reprint,
amsmath,amssymb,
aps,
onecolumn,
]{revtex4-2}
\usepackage{adjustbox}
\usepackage{graphicx}
\usepackage{dcolumn}
\usepackage{bm}
\usepackage{hyperref}
\usepackage[mathlines]{lineno}
\usepackage{caption}
\usepackage{color}
\usepackage{amsthm}

\newtheorem{proposition}{Proposition}

\begin{document}
\title{Genuine Multipartite Nonlocality sharing under Sequential Measurement }
\author{Sk Sahadat Hossain}
\email{sk.shappm2009@gmail.com}
\affiliation{Department of  Mathematics, Nabagram Hiralal Paul College, Hooghly-712246, West Bengal, India}
\author{Indrani Chattopadhyay}
\email{icappmath@caluniv.ac.in}
\affiliation{Department of Applied Mathematics, University of Calcutta, 92, A.P.C. Road, Kolkata-700009, India }

\begin{abstract}
The study of quantum nonlocality sharing has garnered significant attention, particularly for two-qubit and three-qubit entangled systems. In this paper, we extend the investigation to \(n\)-qubit Greenberger-Horne-Zeilinger (GHZ) systems ($ n\geq 4) $, analyzing nonlocality sharing under unbiased unsharp measurements. Employing the Seevink and Svetlichny inequalities, we explore both unilateral, bilateral and possibly multilateral sequential measurement scenarios. In the unilateral scenario, we derive the range of the sharpness parameter ($ \lambda $) for which an observer's multiple copies can share genuine multipartite nonlocality with single copies of the remaining parties. In the bilateral scenario, we identify the maximum number of sequential observers that can share genuine multipartite nonlocality with other parties. A crucial aspect of our results is that all findings stem from a measurement strategy where each sequential observer utilizes unbiased unsharp measurements. As a specific case, for the four-qubit maximally entangled GHZ state, we demonstrate that at most two copies of an observer (e.g., Alice) can share nonlocality in the unilateral sequential measurement scenario. However, in the bilateral scenario, no additional sharing is possible compared to the unilateral case. Consequently, the bilateral result suggests that no further advantage is achieved in the multilateral scenario. This finding highlights the significance of unsharp measurements in optimizing the recycling of qubits for generating quantum nonlocality.  

\end{abstract}

\date{\today}
\maketitle

\section{Introduction}
The concept of nonlocality in quantum mechanics has been a subject of intense study and debate since the seminal work of Einstein, Podolsky, and Rosen in the 1930s \cite{1}. In 1964, John Bell established experimentally testable criteria, known as Bell's inequality \cite{2}. Bell's inequality demonstrates that certain entangled states cannot be expressed in terms of local hidden variable models (LHVs)\cite{2}. This fundamental aspect of quantum mechanics is referred to as quantum nonlocality\cite{3}. 
 In 1969, Clauser, Horne, Shimony, and Holt proposed a simplified form of Bell's inequality, now known as the CHSH inequality \cite{4}. While quantum entanglement \cite{5,6} is a prerequisite for demonstrating Bell nonlocality, it is not sufficient on its own \cite{7}. Nonlocality has been extensively studied for diverse applications \cite{3,8,9,10,11} and across various quantum systems \cite{12,13,14,15}. Nonlocal correlations serve as essential resources for numerous quantum operations, including quantum cryptography \cite{16,17,18}, device-independent quantum key distribution \cite{19,20,21}, randomness generation \cite{22,23,24,25}, and dimension witnessing \cite{26,27}.\\
Recently, considerable attention has been given to the question of whether a single entangled pair can produce a long sequence of nonlocal correlations. For bipartite quantum systems, it was demonstrated in \cite{28,29} that, under specific conditions, at most two Bobs can achieve an expected violation of the Clauser-Horne-Shimony-Holt (CHSH) inequality with a single Alice. This holds when each Bob performs measurements with equal probability and sharpness, and the measurement settings and outcomes of each Bob are independent of those chosen or observed by previous Bobs. In \cite{30}, the authors demonstrated that with unequal sharpness in each Bob's measurements, arbitrarily many independent Bobs can share the nonlocality of the maximally entangled two qubit Bell states with a single Alice. Building on this work, the result was subsequently generalized to higher-dimensional bipartite quantum systems \cite{31}, where the nonlocality of entangled states in higher dimensions was harnessed to enable similar sequential sharing among multiple Bobs. Furthermore, there has been significant progress in extending these ideas to more complex scenarios involving multiple Alices and multiple Bobs interacting with shared quantum resources, \cite{32,33}. These advancements pave the way for understanding the fundamental limits and applications of nonlocal correlations in multipartite and high-dimensional quantum systems. In tripartite systems, the phenomenon of sequential sharing of nonlocality has been extensively investigated and observed in \cite{34,35}. This research demonstrates that nonlocal correlations can be sequentially shared among multiple observers interacting with a shared quantum resource. Furthermore, the sharing of entanglement through sequential measurements by multiple observers has been explicitly demonstrated in the same context \cite{36,37,38}. These studies reveal that successive measurements performed by independent observers on entangled states do not fully destroy the entanglement, allowing it to be utilized by subsequent observers in the sequence. Additionally, it has been shown that multiple observers can sequentially exhibit quantum steering of an entangled state \cite{39,40,41,42}. Quantum steering, a directional form of quantum correlation, allows one party to influence the state of another in a way that cannot be explained by classical means. The sequential demonstration of steering by multiple independent observers highlights the robustness of quantum correlations and further expands the understanding of multipartite and sequential quantum processes. Moreover, the bilateral sharing of nonlocality in two-qubit entangled states \cite{32,43} and the trilateral sharing of nonlocality in three-qubit entangled states \cite{44} have been explored in depth. \\
In a multipartite quantum system, multiple parties share a quantum state, and measurements are performed sequentially on different subsystems. The classification of unilateral, bilateral, and multilateral sequential measurement setups provides a structured framework for analyzing how these measurements are organized. A unilateral sequential measurement setup refers to a scenario in which a single party of a multipartite system conducts measurements on a specific part of the system in a sequential manner. Each measurement is performed step by step, with the outcome of one measurement influencing the subsequent one. A bilateral sequential measurement setup involves two parties (e.g., Alice and Bob) performing sequential measurements on distinct parts of the system. While, the multilateral sequential measurement extends this concept to more than two parties, where multiple observers perform measurements sequentially on different subsystems. The order of these measurements and their inter-dependencies can vary, resulting in a complex flow of information across the system.
 Significant progress has been achieved in advancing the understanding of nonlocality sharing in these and related scenarios, as evidenced by extensive research efforts along this line \cite{45,46,47,48,49}. This capability to share entanglement sequentially broadens the potential applications of quantum resources in practical scenarios, such as quantum communication \cite{50} and distributed quantum computing \cite{51}.\\
It is of considerable importance to investigate whether nonlocality can be shared in unilateral or multilateral configurations involving four or more parties within the framework of the aforementioned formalism \cite{52,53,54}. In this regard, Seevinck and Svetlichny \cite{52}, together with Collins et al. \cite{53}, established sufficient criteria for the detection of genuine \( n \)-party nonlocality. These criteria apply to \( n \)-particle entangled states that cannot be decomposed into convex mixtures of states in which nonlocal correlations are restricted to smaller subsets of particles. In particular, Seevinck and Svetlichny derived \( n \)-party Bell-type inequalities under the assumption of partial separability, where an \( n \)-particle system is considered partially separable if it can be partitioned into subsystems that may exhibit internal correlations, including entanglement, while remaining uncorrelated with one another.
Violations of the inequalities introduced in Refs. \cite{52,53} provide clear evidence of genuine \( n \)-party nonlocality, thereby distinguishing such states from those that are merely partially separable. Notably, these inequalities attain their maximal violations for \( n \)-particle Greenberger–Horne–Zeilinger (GHZ) states, highlighting the exceptional nonlocal character of this class of multipartite entangled states. Furthermore, an explicit and tight upper bound on four-qubit nonlocality has been reported in Ref. \cite{54}, offering valuable insight into the fundamental limits of multipartite nonlocal correlations.
Investigating nonlocality sharing in such multipartite scenarios is essential for understanding the scalability and robustness of nonlocal correlations in complex quantum systems. Moreover, these studies have significant practical implications for emerging quantum technologies, including distributed quantum computing \cite{55}, quantum communication networks \cite{56}, and secure information-processing protocols \cite{57}.\\
This study investigates the distribution of multipartite nonlocality, focusing on systems involving four or more parties, across multiple sequential copies of a given party. Each copy is assumed to perform measurements with uniform probability and identical sharpness, subject to the constraint that the measurement settings and outcomes of each copy are independent of those chosen or obtained by preceding copies. Within this framework, we analyze the multipartite nonlocality formalism introduced by Seevinck and Svetlichny \cite{53} and demonstrate the feasibility of sharing the correlations of multipartite Greenberger–Horne–Zeilinger (GHZ) states among multiple copies in both unilateral and bilateral sequential measurement scenarios for four and higher-qubit systems. In the simplest configuration, consisting of four spatially separated observers \( A_i \) (\( i = 1, \ldots, 4 \)), each equipped with two dichotomous measurement settings, we establish that under specific conditions, at most two copies of \( A_1 \) can simultaneously exhibit genuine four-qubit nonlocality by violating the Seevinck-Svetlichny inequality \cite{52}, while the remaining observers are represented by a single copy in a unilateral sequential measurement setup. Conversely, in the bilateral sequential measurement scenario, no additional advantage is observed when compared to the non-sequential measurement case. This concept has intriguing practical applications across various fields, particularly in quantum communication \cite{58}, quantum metrology \cite{59}, and nonlocality experiments \cite{60}.\\

This article is organized as follows: In section II, we briefly discuss multipartite nonlocality and sequential measurement formalism. In section III, we investigated the unilateral and multilateral sequential sharing of multipartite GHZ state. Finally, we conclude with the conclusion in section IV.
\section{ Basic tools } \label{md}
\subsection{Recapitulation of Multipartite Nonlocality}\label{secI}
In this section, we revisit some fundamental concepts of multipartite nonlocality essential for our study. Consider a system comprising \( n \) particles, each spatially separated in \( n \) distinct directions. Each observer associated with a particle performs two dichotomic measurements on their respective subsystems, where the possible outcomes of the measurements are \( \pm 1 \). The observables associated with these measurements are represented as \( A_1, A_2, \dots, A_n \), corresponding to each of the \( n \) observers. To formalize the scenario, let \( x_i^1, x_i^2 \) denote the inputs (measurement settings) for the \( A_i \)-th observer, where \( x_i^1, x_i^2 \in \{ 0, 1 \} \). The outputs (measurement results) are denoted as \( a_i^1, a_i^2 \), where \( a_i^1, a_i^2 \in \{ 1, -1 \} \). The correlations between the measurement results are described by the joint probability distributions: \( \{ P(a_1 a_2 \dots a_n \mid x_1 x_2 \dots x_n) \}, \) which characterize the probabilities of observing specific outcomes \( a_1, a_2, \dots, a_n \), given the measurement settings \( x_1, x_2, \dots, x_n \).\\
These correlations are considered ``partially separable" if they admit a decomposition in terms of a local hidden variable model \( \lambda \). Specifically, the joint probabilities can be expressed as:
\begin{equation}
P(a_1 a_2 \dots a_n \mid x_1 x_2 \dots x_n) = \int q_{\lambda}(a_1 \dots a_g \mid x_1 x_2 \dots x_g) h_{\lambda}(a_{g+1} \dots a_n \mid x_{g+1} x_{g+2} \dots x_n) \, d\lambda,
\end{equation}
where: \( q_{\lambda} \) and \( h_{\lambda} \) are conditional probability distributions for two disjoint subsets of the \( n \) particles, each dependent on the hidden variable \( \lambda \), \( g \) represents the size of the first subset, and \( d\lambda \) is the probability measure over \( \lambda \). Such a representation reflects the assumption of partial separability, where correlations between subsets of particles can be explained by a local hidden variable shared within each subset. Analogous expressions with different groupings and values of \( g \) describe the correlations for other possible partitions of the particles.\\
The expectation value of the product of the measurement outcomes is given by: \( EX(A_1, A_2, \dots, A_n) = \langle A_1 A_2 \dots A_n \rangle \). Each observer \( A_i \) can choose between two dichotomic observables \( A_i^1 \) and \( A_i^2 \). For simplicity of notation, we write \( EX(i_1 i_2 \dots i_n) \) to represent \( EX(A_1^{i_1}, A_2^{i_2}, \dots, A_n^{i_n}) \), where \( i_k \in \{ 1, 2 \} \) indicates the observable chosen by the \( k \)-th observer. The \( n \)-party inequality, whose violation serves as a sufficient criterion for the full nonseparability of the \( n \)-particle system, is given by \cite{52}:
\begin{equation}\label{ss2}
\vert S_n^\pm \vert \leq 2^{n-1},
\end{equation}
where: \( S_n^\pm = \sum_I v_{t(I)}^\pm EX(i_1 i_2 \dots i_n) \), \( v_{t(I)}^\pm = (-1)^{\frac{t(I)(t(I) \pm 1)}{2}} \), \( I = (i_1 i_2 \dots i_n) \) is a multi-index representing the choice of observables, and \( t(I) \) is the number of times the index \( 2 \) appears in \( I \). These sequences of sign in $v_{t(I)}^\pm  $ have period four with cycles $ (1,-1,-1,1) $ and $ (1,1,-1,-1) $. We call these inequalities alternating.
This inequality applies to the expectation values of the observables and serves as a generalization of the well-known Bell inequalities to multipartite systems. A violation of this inequality indicates that the system cannot be described by any local hidden variable model, thereby confirming the full nonseparability of the \( n \)-particle system. For a detailed derivation and discussion of the coefficients \( v_{t(I)}^\pm \) and their role in the inequality, as well as applications to specific quantum states, refer to Refs. \cite{52,53}.\\
For, $ n=5 $ the inequality takes the explicit form,\\
\begin{equation}
\begin{aligned}
 & \vert
   EX(11111) + EX(21111) + EX(12111) + EX(11211) + EX(11121) + EX(11112)- EX(22111) - EX(21211)\\
 &\quad  - EX(21121) - EX(21112) - EX(12211) - EX(12121)- EX(12112) - EX(11221)\\
&\quad   - EX(11212) - EX(11122) - EX(22211) - EX(22121)- EX(22112) - EX(21221)  \\
&\quad  - EX(21212)- EX(21122) - EX(12221) - EX(12212) - EX(12122) - EX(11222)\\
&\quad  + EX(22221) + EX(12222) + EX(21222) + EX(22122)+ EX(22212) + EX(22222)\vert \leq 16.
\end{aligned}
\end{equation}
This inequality is maximally violated by five qubit GHZ state.
\subsection{ Sequential Measurement Scenario with Multiple Observers }\label{secII}
Silva et al. \cite{28} introduced a groundbreaking perspective on quantum measurement scenarios, which we build upon in this study. By utilizing the foundational principles of their work and incorporating the methodologies proposed in \cite{29,61}, we develop a comprehensive framework for exploring quantum measurements involving multiple observers. For clarity and simplicity, we focus initially on a four-party system involving the following spatially separated participants: \( A_1, A_2, A_3, \) and \( A_4 \). In this configuration, the first party, \( A_1 \), hosts a sequence of observers referred to collectively as ``Alices" (\( \text{Alice}^1 \rightarrow \text{Alice}^2 \rightarrow \text{Alice}^3 \rightarrow \dots \rightarrow \text{Alice}^r \)). These party's share a system of four spin-\( \frac{1}{2} \) particles prepared in the state \( \rho \). Consider, Alices (\( \text{Alice}^1, \text{Alice}^2, \dots, \text{Alice}^r \)): Alices sequentially perform measurements on the first particle. After completing their measurement, each Alice passes the post-measurement state to the next Alice in the sequence, ensuring that the first particle is measured multiple times in a sequential manner. \( A_2 \) performs measurements on the second particle, \( A_3 \) measurements on the third particle, and measurements on the fourth particle are performed by \( A_4 \). The described framework can be generalized to an \( n \)-partite system. In such systems, one-sided sequential sharing of nonlocal correlations is achieved by assigning a sequence of observers to one of the \( n \) parties. For example, in an \( n \)-partite quantum system, one party (e.g., \( A_1 \)) may host a sequence of observers who measure the particle assigned to that party in a sequential manner, while the remaining parties \( A_2, A_3, \dots, A_n \) each perform independent measurements on their respective particles.\\
\textbf{ Bilateral Sequential Sharing of Nonlocal Correlations:} In a more sophisticated extension, the framework allows for bilateral sequential sharing, where sequences of observers are assigned to two or more parties in an \( n \)-partite quantum system. For instance, in a four-party system: a sequence of Alices (\( \text{Alice}^1, \text{Alice}^2, \dots, \text{Alice}^r \)) can be assigned to \( A_1 \). Simultaneously, another sequence of Bobs (\( \text{Bob}^1, \text{Bob}^2, \dots, \text{Bob}^s \)) can be assigned to \( A_2 \). In this scenario, both sequences of observers independently perform measurements on their respective particles sequentially. This bilateral approach enables the study of complex scenario in the sharing and distribution of quantum nonlocal correlations across multiple observers and particles.\\
In the sequential measurement framework, the process begins with \( \text{Alice}^1 \) measuring the first particle. Upon completing their measurement, \( \text{Alice}^1 \) delivers the resulting post-measurement state to \( \text{Alice}^2 \), who subsequently performs their measurement. This sequential process continues, with each observer passing the particle along to the next, until the final observer, \( \text{Alice}^r \), conducts their measurement. A fundamental assumption in this setup is that each Alice operates independently of the measurement decisions and outcomes of the preceding Alices. This independence ensures that the measurements made by any \( \text{Alice}^m \) are unbiased with respect to the choices of \( \text{Alice}^{m-1} \) or any earlier observers in the sequence. Consequently, the system reflects a scenario where each observer acts without influence from the measurement settings or results of others. To further simplify the analysis, an unbiased input scenario is assumed. Under this assumption, all possible measurement choices of each Alice are equally likely, implying a uniform distribution of measurement settings across all observers. This ensures that no specific measurement choice is given preferential consideration, facilitating a balanced exploration of the system.
An essential aspect of multipartite quantum systems is the no-signaling condition, which mandates that the measurement outcomes of one party should not depend on the measurement settings of spatially separated parties. In the described setup, this condition is satisfied between \( A_1, A_2, A_3, \dots, A_n \), as these observers are spatially separated and perform measurements on distinct particles. This spatial separation ensures that information about the measurement settings or outcomes of one observer cannot influence those of another. However, the no-signaling condition does not hold between the sequential Alices in \( A_1 \). Since each Alice in the sequence measures the same particle, the measurement decisions and outcomes of \( \text{Alice}^{m-1} \) directly influence the state of the particle passed to \( \text{Alice}^m \). As a result, \( \text{Alice}^{m-1} \) effectively signals to \( \text{Alice}^m \) through their choice of measurements, where \( m \in \{2, 3, \dots, r\} \). This signaling arises because the particle's state, as received by \( \text{Alice}^m \), depends on the specific operations performed by \( \text{Alice}^{m-1} \). This formulation highlights the interplay between spatial separation, no-signaling constraints, and sequential measurement formalism in multipartite quantum systems.\\
In the present sequential measurement framework under consideration, each observer in the sequence of Alices, except for the final observer \( \text{Alice}^r \), is constrained in their ability to perform sharp measurements on the particle under their control. If any of the first \( (m-1) \) Alices performs a sharp measurement on the particle, the quantum entanglement among the remaining parties (\( \text{Alice}^{m}, \text{Alice}^{m+1}, \dots, \text{Alice}^r, A_2, A_3, \dots, A_n \)) will be irreversibly destroyed. The sharp measurement collapses the quantum state, severing the entanglement required for maintaining quantum correlations across the system. This destruction of entanglement renders subsequent observers incapable of achieving violations of the multipartite inequalities, such as those presented in Eq. \eqref{ss2}. Consequently, the ability of the system to demonstrate quantum nonlocality is effectively nullified for the remaining observers. To mitigate the detrimental effects of sharp measurements and ensure the preservation of entanglement, the measurements performed by the first \( (r-1) \) Alices must be weak. Weak measurements are characterized by their minimal disturbance to the quantum state, allowing for the sequential sharing of quantum correlations without fully collapsing the state. This careful balancing of information extraction and state preservation is critical for enabling meaningful measurement outcomes across all observers. In this study, we employ the weak measurement formalism as developed in \cite{28}, utilizing its unsharp variant as described in \cite{29,41}. Unsharp measurements allow for a controlled degree of measurement sharpness, parameterized to achieve an optimal trade-off between information gain and the preservation of quantum correlations. This formalism ensures that the sequential measurements can proceed without compromising the system's ability to exhibit multipartite nonlocality. In this context, it is noteworthy that the sequential sharing of nonlocality under projective measurement has been recently demonstrated, as reported in \cite{62,63,64,65}. \\

In the context of standard von Neumann measurements, the state of a spin-\( \frac{1}{2} \) particle, initially described by \( \vert \psi \rangle = a \vert 0 \rangle + b \vert 1 \rangle \), evolves following its interaction with a measurement apparatus (meter) in the state \( \vert \varphi(q)\rangle \). Here, \( \vert 0 \rangle \) and \( \vert 1 \rangle \) form an orthonormal basis in \( \mathbb{C}^2 \), and the normalization condition \( \vert a \vert^2 + \vert b \vert^2 = 1 \) holds. After the interaction, the joint state of the particle and the meter becomes:  
\begin{equation}
a \vert 0 \rangle \otimes \vert \varphi(q-1)\rangle + b \vert 1 \rangle \otimes \vert \varphi(q+1)\rangle.
\end{equation}
This expression reflects the coupling between the particle's spin state and the meter's position state, with the meter's position \( q \) shifted based on the spin state of the particle (for notational simplicity, we henceforth denote this state by $ \varphi(q) $ only)).\\
In Ref. \cite{28}, the weak measurement formalism was redefined without relying on post-selection strategies \cite{66}, introducing two key parameters: the quality factor \( F \) and the precision \( G \) of measurements. The quality factor, defined as \( F(\varphi) = \int_{-\infty}^{\infty} \varphi (q+1) \varphi (q-1) dq, \) quantifies the extent to which the post-measurement state remains proportional to the original state. Conversely, the precision factor, defined as \( G(\varphi) = \int_{-1}^{1} \varphi^{2}(q) dq, \) measures the amount of information gained from the measurements. For strong projective measurements, \( F = 0 \) and \( G = 1 \). An optimal pointer state achieves the best trade-off between these two factors, maximizing precision for a given quality factor. It has been demonstrated that for such optimal pointer states, the information gain-disturbance trade-off satisfies the relation \( F^2 + G^2 = 1 \) \cite{28}.\\
This weak measurement framework has been reformulated within the unsharp measurement formalism in \cite{29,41}, which represents a specific class of Positive Operator-Valued Measurements (POVMs) \cite{61,67}. A POVM is defined as a set of positive operators that sum to the identity operator:  
\(
\mathcal{E} \equiv \{E_i \mid \Sigma_i E_i = I, \; 0 < E_i \leqslant I \},  
\)  
where the effect operators \( E_i \) correspond to quantum events associated with possible measurement outcomes. For a dichotomic unsharp measurement scenario, the effect operators are defined as  
\(
E_{\pm}^\lambda = \frac{I_2 \pm \lambda \widehat{n} \cdot \vec{\sigma}}{2},  
\) where \( \lambda \) is the sharpness parameter (\( 0 < \lambda \leq 1 \)), \( \vec{\sigma} = (\sigma_1, \sigma_2, \sigma_3) \) represents the vector of Pauli matrices, \( \widehat{n} \) is a unit vector in three dimensions, and \( I_2 \) is the \( 2 \times 2 \) identity matrix. The probabilities of obtaining outcomes \( +1 \) and \( -1 \) are given by \( \mathrm{Tr}[\rho E_+^\lambda] \) and \( \mathrm{Tr}[\rho E_-^\lambda] \), respectively, where \( \rho \) is the quantum state under consideration. The post-measurement states are determined by the von Neumann-Luders transformation rule \cite{61,63} and are expressed as \(
\frac{\sqrt{E_+^\lambda} \rho \sqrt{E_+^\lambda}}{\mathrm{Tr}[\rho E_+^\lambda]} \quad \text{and} \quad \frac{\sqrt{E_-^\lambda} \rho \sqrt{E_-^\lambda}}{\mathrm{Tr}[\rho E_-^\lambda]},  
\) corresponding to the \( +1 \) and \( -1 \) outcomes, respectively, up to local unitary freedom. It is well established \cite{29} that the weak measurement parameters \( F \) and \( G \) introduced in \cite{28} are directly linked to the unsharp measurement formalism through the relations \( F = \sqrt{1 - \lambda^2} \) and \( G = \lambda \). Thus, \( \lambda = 1 \) corresponds to strong projective measurements (\( G = 1, F = 0 \)), and the sharpness parameter \( \lambda \) serves as an indicator of the measurement's precision. Additionally, the optimal pointer state condition \( F^2 + G^2 = 1 \) is inherently satisfied within this framework. For the purposes of this work, we will assume that all Alices, except for the final observer, perform unsharp measurements to maintain the entanglement of the quantum system and enable subsequent analyses.
\section{Sharing of genuine multipartite nonlocality by multiple Alices}
After introducing the foundational concepts necessary for our study, we turn to the central question: how many Alices can share multipartite nonlocality with the remaining parties \( A_2, A_3, \dots, A_n \) through the violation of the inequality \eqref{ss2}? To simplify notation and enhance clarity, we denote the parties that are not performing sequential measurements as \( B_{j} \), where \( j \in \{2, 3, \dots, n\} \), and the multiple Alice observers as Alice\( ^{1} \), Alice\( ^{2} \), \(\dots\), Alice\( ^{r} \). The measurement outcomes corresponding to multiple Alice observers are denoted as \( a^{m} \in \{1,-1\} \), where \( m = 1,2, \dots, r \), while the measurement outcomes for the observable \( B_{j} \) are represented as \( b_{j} \in \{1,-1\} \). To explore this, we assume that multiple Alices, along with \( B_2, B_3, \dots, B_n \), initially share the \( n \)-qubit GHZ state. The state is expressed as 
\begin{equation}\label{dd1}
\rho_{\text{GHZ}} = \vert \Phi \rangle \langle \Phi \vert,   
\end{equation}
where  
\[
\vert \Phi \rangle = \frac{1}{\sqrt{2}} (\vert 00 \ldots 0 \rangle + \vert 11 \ldots 1 \rangle).
\]  
Here, \( \lbrace \vert 0 \rangle, \vert 1 \rangle \rbrace \) form an orthonormal basis in the two-dimensional Hilbert space \( \mathbf{H} \). The GHZ state serves as a maximally entangled state for \( n \)-partite systems, making it ideal for studying the sharing of nonlocality in multipartite quantum systems. We consider that the measurement directions for the spin component observables of \( \text{Alice}^m \) (where \( m \in \lbrace 1, 2, \dots, r \rbrace \)) are denoted by \( \lbrace \widehat{x}_0^m, \widehat{x}_1^m \rbrace \). Similarly, the measurement directions for the remaining parties \( B_2, B_3, \dots, B_n \) are given by \( \lbrace \widehat{y_{0}}^j, \widehat{y_{1}}^j \rbrace \) for \( j = 2, 3, \dots, n \). The measurement outcomes for these settings are binary and take values \( \pm 1 \). Let us assume, that the two possible choices of the measurement settings of Alice$^{m}$ are the spin component observables in the direction $ \widehat{x^{m}} $, where
\begin{equation}\label{dd2}
\widehat{x_{i}^{m}}= \sin\theta_{i}^{x^{m}}\cos\phi_{i}^{x^{m}} \widehat{X}+\sin\theta_{i}^{x^{m}}\sin\phi_{i}^{x^{m}} \widehat{Y}+\cos\theta_{i}^{x^{m}} \widehat{Z},
\end{equation}
where $ 0\leq \theta_{i}^{x^{m}}\leq \pi $;  $ 0\leq \phi_{i}^{x^{m}}\leq 2\pi $, with $ i \in \lbrace 0,1\rbrace $, while $ \widehat{X},\;\widehat{Y}$ and $ \widehat{Z} $ are three mutually orthogonal unit vectors in the Cartesian coordinate system.\\
In the Bilateral sequential sharing scenario, the measurement directions for the spin component observables of \( \text{Bob}^k \) (where \( k \in \lbrace 1, 2, \dots, s \rbrace \)) are denoted by \( \lbrace \widehat{\overline{x}}_0^k, \widehat{\overline{x}}_1^k \rbrace \), where
\begin{equation}\label{ddj2}
\widehat{\overline{x}_{i}^{k}}= \sin\theta_{i}^{\overline{x}^{k}}\cos\phi_{i}^{\overline{x}^{k}} \widehat{X}+\sin\theta_{i}^{\overline{x}^{k}}\sin\phi_{i}^{\overline{x}^{k}} \widehat{Y}+\cos\theta_{i}^{\overline{x}^{k}} \widehat{Z},
\end{equation}
where $ 0\leq \theta_{i}^{\overline{x}^{k}}\leq \pi $;  $ 0\leq \phi_{i}^{\overline{x}^{k}}\leq 2\pi $, with $ i \in \lbrace 0,1\rbrace $.\\
Similarly, the two possible choices of the measurement settings of B$_{j}$ are the spin component observables in the direction $ \widehat{y^{j}} $, where
\begin{equation}\label{rdd2}
\widehat{y_i^{j}}= \sin\theta_{i}^{y_j}\cos\phi_{i}^{y_j} \widehat{X}+\sin\theta_{i}^{y_j}\sin\phi_{i}^{y_j} \widehat{Y}+\cos\theta_{i}^{y_j} \widehat{Z},
\end{equation}
where $ 0\leq \theta_{i}^{y_j}\leq \pi $;  $ 0\leq \phi_{i}^{y_j}\leq 2\pi $, with $ i \in \lbrace 0,1\rbrace $, while $ \widehat{X},\;\widehat{Y}$ and $ \widehat{Z} $ are three mutually orthogonal unit vectors in the Cartesian coordinate system ( for simplicity from here onward we will use $x^{m},$  $ \overline{x}^{k} $ and $ y^{j} $ as measurement directions, respectively).\\
The aim of the study is to determine the maximum number of Alices who can perform sequential measurements on the first particle of the GHZ state while still allowing the multipartite nonlocality of the state to be shared with the remaining parties \( B_2, B_3, \dots, B_n \). This analysis involves evaluating the interplay between the weak measurement formalism, the entanglement properties of the GHZ state, and the violation of the multipartite inequality \eqref{ss2}. \\
The choice of measurement settings and the sequential measurement protocol by the Alices are key factors in ensuring that nonlocal correlations are preserved, thereby enabling a violation of the inequality. The study also examines how the weak measurement parameters, particularly the sharpness parameter \( \lambda \), influence the ability to share nonlocality across multiple Alices and the remaining parties. In subsequent sections, we will calculate and analyze the multipartite inequality violation under this framework, considering both the quality and precision of measurements and the limitations imposed by the sequential measurement process.\\
The conditional probabilities for the outcomes \( a^{1}, a^{2}, b_2, \ldots, b_n \) (with \( a^{i}, b_j \in \lbrace 1, -1 \rbrace \)) are defined as\\
 $ P(a^{1}, a^{2}, b_2, \ldots, b_n \vert x^1, x^2, y^{2}, \ldots, y^{n}),$ given that \( \text{Alice}^1, \text{Alice}^2 \), and the parties \( B_2, B_3, \ldots, B_n \) perform measurements of spin component observables in the directions \( x^1, x^2, y^2, y^3\ldots, y^n \), respectively. Here, \( \lambda_m \) represents the sharpness parameter for \( \text{Alice}^m \) (where \( m \in \lbrace 1, 2, 3, \ldots, r-1 \rbrace \)), and the last observer, \( \text{Alice}^r \), always performs sharp measurements with \( \lambda_r = 1 \). The conditional probabilities are determined using the Born rule as follows:  
\begin{equation}\label{dd4}
P(a^{1}, a^{2}, b_2, \ldots, b_n \vert x^1, x^2, y^{2},y^3 \ldots, y^{n}) = \operatorname{Tr}[ E_{a^{2}}^{\lambda_2} \cdot \rho_{\text{un}}^{A^2}]. 
\end{equation}
Here, the effect operator \( E_{a^2}^{\lambda_2} \) is given by \(
E_{a^2}^{\lambda_2} = \lambda_2 \frac{ I_2 + a^2 \, \widehat{x^2} \cdot \vec{\sigma}}{2} + (1 - \lambda_2) \frac{I_2}{2}, \) with $ \lambda_2 = 1 $ in this scenario. Here, \( \vec{\sigma} = (\sigma_1, \sigma_2, \sigma_3) \) denotes the vector of Pauli matrices, and \( I_2 \) is the \( 2 \times 2 \) identity matrix. The term \( \rho_{\text{un}}^{A^2} \) represents the unnormalized quantum state at the side of \( \text{Alice}^2 \) (\( A^2 \)), conditioned on the measurements performed by \( \text{Alice}^1 \) and the parties \( B_2, B_3, \ldots, B_n \) on their respective spin components along directions \( x^1, y^2, y^3 \ldots, y^n \), with outcomes \( a^1, b_2, \ldots, b_n \).\\
Consequently, in a scenario with $ r $ Alice the conditional probability are given by:
\begin{equation}\label{pdd5}
 P(a^{1}, a^{2},\ldots a^r, b_2, \ldots, b_n \vert x^1, x^2,\ldots x^r, y^{2},y^3 \ldots, y^{n}) = \operatorname{Tr}[ E_{a^{r}}^{\lambda_r} \cdot \rho_{\text{un}}^{A^r}], 
 \end{equation} 
 where, the effect operator \( E_{a^r}^{\lambda_r} \) is given by \(
E_{a^r}^{\lambda_r} = \lambda_r \frac{ I_2 + a^r \, \widehat{x^r} \cdot \vec{\sigma}}{2} + (1 - \lambda_r) \frac{I_2}{2}, \) with $ \lambda_r = 1 $ as Alice$ ^{r} $ perform sharp measurements. The term \( \rho_{\text{un}}^{A^r} \) represents the unnormalized quantum state at the side of \( \text{Alice}^r \) (\( A^r \)), conditioned on the measurements performed by \( \text{Alice}^1, \text{Alice}^2,\ldots \text{Alice}^{r-1} \) and the parties \( B_2, B_3, \ldots, B_n \) on their respective spin components along directions \( x^1, x^2,\ldots, x^{r-1}, y^2, y^3 \ldots, y^n \), with outcomes \( a^1, a^2, \ldots a^{r-1}, b_2, \ldots, b_n \). This formulation encapsulates the interplay between the unsharp measurement formalism, characterized by the sharpness parameter \( \lambda_m \), and the evolution of the quantum state as it is sequentially measured by multiple observers. The unnormalized state \( \rho_{\text{un}}^{A^2} \) is expressed as follows:  
\begin{equation}\label{dd5}
\rho_{\text{un}}^{A^2} = \operatorname{Tr}_{B_2B_3\ldots B_n} \Big[ \Big\{ \sqrt{E_{a^1}^{\lambda_1}} \otimes \frac{I_2 + b_2 \, \widehat{y^2} \cdot \vec{\sigma}}{2} \otimes \cdots \otimes \frac{I_2 + b_n \, \widehat{y^n} \cdot \vec{\sigma}}{2} \Big\} \cdot \rho_{GHZ} \cdot \Big\{ \sqrt{E_{a^1}^{\lambda_1}} \otimes \frac{I_2 + b_2 \, \widehat{y^2} \cdot \vec{\sigma}}{2} \otimes \cdots \otimes \frac{I_2 + b_n \, \widehat{y^n}_n \cdot \vec{\sigma}}{2} \Big\} \Big],   
\end{equation}
where the square root of the effect operator \( E_{a^1}^{\lambda_1} \) is given by:  
\[
\sqrt{E_{a^1}^{\lambda_1}} = \sqrt{\frac{1 + \lambda_1}{2}} \cdot \frac{I_2 + a^1 \, \widehat{x^1} \cdot \vec{\sigma}}{2} + \sqrt{\frac{1 - \lambda_1}{2}} \cdot \frac{I_2 - a^1 \, \widehat{x^1} \cdot \vec{\sigma}}{2}.  
\]  
Here:  \( \vec{\sigma} = (\sigma_1, \sigma_2, \sigma_3) \) represents the vector of Pauli matrices. \( \widehat{x^1} \) and \( \widehat{y^j} \) are unit vectors indicating the measurement directions of \( \text{Alice}^1 \) and the parties \( B_2, B_3, \ldots, B_n \), respectively. \( \lambda_1 \) is the sharpness parameter associated with the measurement performed by \( \text{Alice}^1 \). \( \rho_{GHZ} \) is the initial GHZ state shared by the observers. \( \operatorname{Tr}_{B_2B_3\ldots B_n} \) denotes the partial trace over the subsystems \( B_2, B_3, \ldots, B_n \). This formulation captures the evolution of the quantum state under sequential unsharp measurements performed by the parties involved. \\
The joint probability of outcome $ P(a^2, b_2, \ldots, b_n \vert x^1, x^2, y_2, \ldots, y_n) $ are given by
\begin{equation}\label{dd8}
P(a^2, b_2, \ldots, b_n \vert x^1, x^2, y_2, \ldots, y_n) =  \Sigma_{a^1=+1,-1} P(a^1, a^2, b_2, \ldots, b_n \vert x^1, x^2, y_2, \ldots, y_n).
\end{equation}
The correlation \( EX^2(i^1 i^2 j_2 \ldots j_n) \) ( where $ i^1,i^2, j_l \in \lbrace 1,2\rbrace,$ and $ l= 2,3, \ldots n $) among \( \text{Alice}^2 \) and \( B_2, B_3, \ldots, B_n \) is determined as follows:  
\begin{equation}\label{dd9}
EX^2(i^1 i^2 j_2 \ldots j_n) = \sum_{a^2, b_2, \ldots, b_n = \pm 1} (a^2 b_2 \cdots b_n) P(a^2, b_2, \ldots, b_n \vert x^1, x^2, y_2, \ldots, y_n).
\end{equation}
Since \( \text{Alice}^2 \) does not have knowledge of the measurement directions chosen by \( \text{Alice}^1 \), the correlation given in Eq.~\eqref{dd9} is averaged over the two possible measurement settings of \( \text{Alice}^1 \). Consequently, the average correlations among \( \text{Alice}^2 \) and \( B_2, B_3, \ldots, B_n \) are expressed as:  
\begin{equation}\label{dd10}
\overline{EX^2(i^2 j_2 \ldots j_n)} = \sum_{i^1 = 1, 2} EX^2(i^1 i^2 j_2 \ldots j_n) P(x^1)),
\end{equation}
where \( P(x^1) \) denotes the probability of \( \text{Alice}^1 \)'s spin component measurement in the direction \( x^1 \). For an unbiased measurement setting, the probability of each measurement direction is equal, i.e., \( P(x^1)) = \frac{1}{2} \). Substituting this into Eq.~\eqref{dd10} gives:  
\begin{equation}
\overline{EX^2(i^2 j_2 \ldots j_n)} = \frac{1}{2} \sum_{i^1 = 1, 2} EX^2(i^1 i^2 j_2 \ldots j_n).
\end{equation}
As a result, the modified \( n \)-qubit inequality for the correlations among \( \text{Alice}^2 \) and \( B_2, B_3, \ldots, B_n \) can be formulated as:  
\begin{equation}\label{dd11}
\vert \overline{S_n^\pm} \vert \leq 2^{n-1}.
\end{equation}
This inequality establishes the conditions for sharing multipartite nonlocality and serves as the benchmark for determining the maximum number of Alices who can share nonlocal correlations with the remaining parties.
\begin{figure}[htbp]
\centering
\includegraphics[width=0.6\linewidth,trim=1cm 1cm 1cm 1cm, clip]{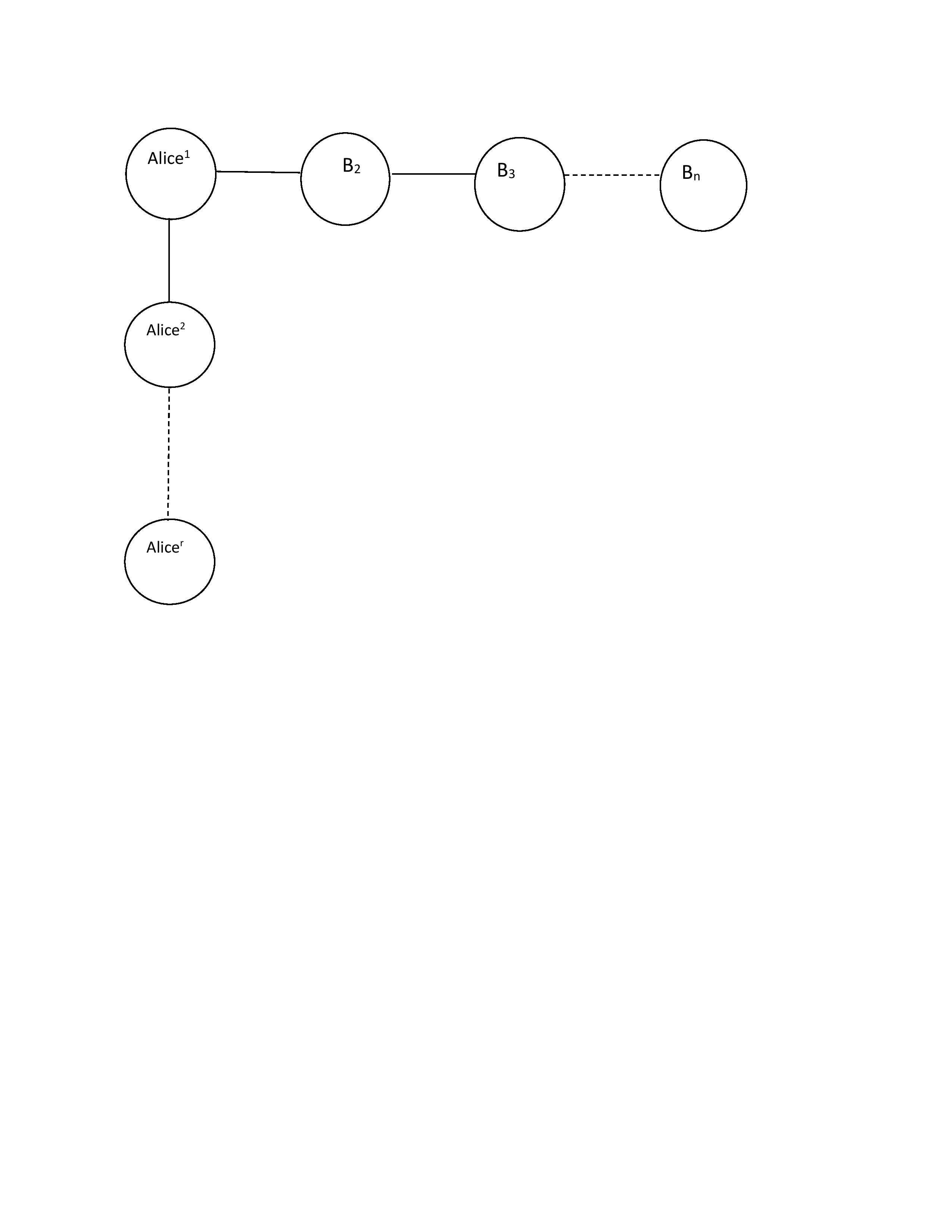}
\captionsetup{skip=-8pt,belowskip=0pt}
\caption{ Sharing of unilateral multipartite nonlocal correlation under sequential measurement. In this figure, in an n-partite system, multiple Alices are exhibiting nonlocality with rest of the parties.}
\label{sd1}
\end{figure}\\
\subsection{ Unilateral Sharing of Multipartite Nonlocality}\label{sec3}
We investigate the simultaneous sharing of multipartite nonlocality when multiple observers (Alice's) sequentially measure their respective parts of a quantum system. Specifically, we focus on the simplest multipartite scenario involving four qubits and extend our findings to \( n \)-qubit systems. Consider a four-qubit system, where Alice\(^1\) and $ B_2, B_3, B_4 $ performs measurements on their respective copies of the GHZ state.\\
We show that, with a proper choice of measurement settings, the expectation values among Alice$ ^{1} $ and the remaining parties in the four-qubit system can be obtained as 
\begin{equation}
 \overline{EX(i^{1}j_{2}j_{3}j_{4})}=\lambda_{1} \cos(\phi_{i}^{x^{1}}+\phi_{j}^{y_2}+\phi_{j}^{y_3}+\phi_{j}^{y_4})
\end{equation} 
 The optimized correlations of the corresponding Bell-type inequality is given by:
\begin{equation}\label{dd12} 
S_4^1 = 8\sqrt{2} \lambda_1,
\end{equation}
where \( \lambda_1 \) is the sharpness parameter associated with the measurement settings of Alice\(^1\). The measurement settings for each observer are chosen as follows: Alice$ ^1 $ : $ (\theta_{1}^{x^1},\;\phi_{1}^{x^1},\;\theta_{2}^{x^1},\;\phi_{2}^{x^1})\; \equiv \;(\dfrac{\pi}{2},\;0,\; \dfrac{\pi}{2},\;\dfrac{\pi}{2}) $,  B$_2 $ : $ (\theta_{1}^{y_2},\;\phi_{1}^{y_2},\;\theta_{2}^{y_2},\;\phi_{2}^{y_2})\; \equiv \;(\dfrac{\pi}{2},\;\dfrac{\pi}{5},\; \dfrac{\pi}{2},\;\dfrac{7 \pi}{10}) $, B$_3 $ : $ (\theta_{1}^{y_3},\;\phi_{1}^{y_3},\;\theta_{2}^{y_3},\;\phi_{2}^{y_3})\; \equiv \;(\dfrac{\pi}{2},\;\dfrac{8\pi}{5},\; \dfrac{\pi}{2},\;\dfrac{\pi}{10}),$ and B$ _4 $: $ (\theta_{1}^{y_4},\;\phi_{1}^{y_4},\;\theta_{2}^{y_4},\;\phi_{2}^{y_4})\; \equiv \;(\dfrac{2\pi}{5},\;\dfrac{8\pi}{5},\; \dfrac{\pi}{2},\;\dfrac{9\pi}{10}),$.\\
After Alice\( ^1 \) performs an unsharp measurement, she transmits the post-measurement state to Alice\( ^2 \), who subsequently performs measurements on the received state. The measurement settings for Alice\( ^2 \) are: $ (\theta_{1}^{x^2},\;\phi_{1}^{x^2})$ and $\theta_{2}^{x^2},\;\phi_{2}^{x^2}) $. 
We find that, upon an appropriate selection of measurement settings, the expectation values corresponding to Alice$^{2}$ and the remaining parties in the four-qubit system take the form
\begin{equation}
 \overline{EX^2(i^2j_{2}j_{3}j_{4})}=\frac{\lambda_{2}}{2} (1+\sqrt{1-(\lambda_1)^2})\cos(\phi_{i}^{x^{2}}+\phi_{j}^{y_2}+\phi_{j}^{y_3}+\phi_{j}^{y_4})
\end{equation} 
Optimizing over the measurement settings of Alice\( ^2 \), the nonlocality parameter for the correlations among Alice\( ^2 \) and the remaining parties is given by:  
\begin{equation}\label{dd13}  
S_4^2 = 4\sqrt{2} \lambda_2 \left( 1 + \sqrt{1 - (\lambda_1)^2} \right),
\end{equation}  
where \( \lambda_2 \) represents the sharpness parameter associated with the measurements performed by Alice\( ^2 \). This expression suggests that both Alice\( ^1 \) and Alice\( ^2 \) can simultaneously demonstrate four-qubit nonlocality with the other parties, provided that \( \lambda_1 > \frac{1}{\sqrt{2}} \). \\
Following her measurement, Alice\( ^2 \) transmits the post-measurement state to Alice\( ^3 \), who then performs a measurement on the received state. 
The expectation values corresponding to correlations between Alice$^{3}$ and the remaining non-sequential parties are given by
\begin{equation}
 \overline{EX^3(i^3j_{2}j_{3}j_{4})}=\frac{\lambda_{3}}{4} (1+\sqrt{1-(\lambda_1)^2}+\sqrt{1-(\lambda_2)^2}+\sqrt{1-(\lambda_1)^2}\sqrt{1-(\lambda_2)^2})\cos(\phi_{i}^{x^{3}}+\phi_{j}^{y_2}+\phi_{j}^{y_3}+\phi_{j}^{y_4})
\end{equation} 
The nonlocality parameter associated with the measurements performed by Alice\( ^3 \) and the remaining parties is given by: 
\begin{equation}\label{dd14}  
S_4^3 = 2\sqrt{2} \lambda_3 \left( 1 + \sqrt{1 - (\lambda_1)^2} + \sqrt{1 - (\lambda_2)^2} + \sqrt{1 - (\lambda_1)^2} \sqrt{1 - (\lambda_2)^2} \right),
\end{equation}  
where \( \lambda_3 \) denotes the sharpness parameter for Alice\( ^3 \). \\
For successful sequential sharing of nonlocality among multiple Alice observers, both Alice\( ^1 \) and Alice\( ^2 \) must perform unsharp measurements to preserve nonlocal correlations for subsequent observers. By imposing the conditions \( \lambda_1 > \frac{1}{\sqrt{2}} \) and \( \lambda_2 > 0.83 \), we obtain:  

\[
S_4^3 = 7.52 \lambda_3.
\]  

Since \( S_4^3 \) does not exceed the threshold required for nonlocality violation, it follows that Alice\( ^1 \), Alice\( ^2 \), and Alice\( ^3 \) cannot simultaneously exhibit four-qubit nonlocality with a single copy of the remaining parties.\\
\textbf{Nonlocality Sharing in a five-qubit System:} For the five-qubit scenario involving multiple Alices and the parties $B_2, B_3, \ldots, B_5$. The expectation values among Alice$ ^{1} $ and the remaining parties in the five-qubit system is given by 
\begin{equation}
	\overline{EX(i^{1}j_{2}j_{3}j_{4}j_{5})}=\lambda_{1} \cos(\phi_{i}^{x^{1}}+\phi_{j}^{y_2}+\phi_{j}^{y_3}+\phi_{j}^{y_4}+\phi_{j}^{y_5})
\end{equation} 
 The correlation observed between Alice$^{1}$ and the remaining parties are given by:
\begin{equation}
S_5^1 = 16\sqrt{2} \lambda_1.
\end{equation}
The expectation values corresponding to Alice$^{2}$ and the remaining parties in the five-qubit system take the form
\begin{equation}
	\overline{EX^2(i^2j_{2}j_{3}j_{4}j_{5})}=\frac{\lambda_{2}}{2} (1+\sqrt{1-(\lambda_1)^2})\cos(\phi_{i}^{x^{2}}+\phi_{j}^{y_2}+\phi_{j}^{y_3}+\phi_{j}^{y_4}+\phi_{j}^{y_5})
\end{equation} 
Under the sequential measurement framework, the correlations between Alice$^{2}$ and the remaining parties are expressed as:
\begin{equation}
S_5^2 = 8\sqrt{2} \lambda_2 \left( 1 + \sqrt{1 - \lambda_1^{2}} \right).
\end{equation}
Within this setting, both Alice$^{1}$ and Alice$^{2}$ are capable of simultaneously exhibiting five-qubit nonlocality provided that $\lambda_1 > \frac{1}{\sqrt{2}}$. However, it is found that Alice$^{1}$, Alice$^{2}$, and Alice$^{3}$ cannot simultaneously demonstrate five-qubit nonlocality with the remaining parties.\\
\textbf{Nonlocality sharing for six-qubit system:} We have found that, for six-qubit system, consisting of multiple Alices and the parties \( B_2, B_3, \ldots, B_n \), the correlations among Alice$ ^{1} $ and the remaining parties are given by:
\begin{equation}
 S_6^1 = 32\sqrt{2} \lambda_1.
 \end{equation} 
The correlations between Alice$ ^{2} $ and the other parties in a sequential measurement scenario are provided by: 
\begin{equation}
S_6^2 = 16\sqrt{2} \lambda_2 \left( 1 + \sqrt{1 - (\lambda_1)^2} \right).
\end{equation}
We find that, Alice$ ^{1} $ and Alice$ ^{2} $ can exhibit six-qubit nonlocality simultaneously for $ \lambda_1 >\frac{1}{\sqrt{2}} $. In contrast, Alice$ ^{1} $, Alice$ ^{2} $ and Alice$ ^{3} $ fails to exhibit six-qubit nonlocality simultaneously with the remaining parties. \\ 
 This consistency across different system sizes strongly suggests that the limitation on simultaneous nonlocality sharing extends to a general \( n \)-qubit GHZ state. Hence, we conclude that, in an \( n \)-qubit system, no more than two Alice observers can share nonlocality simultaneously in a sequential measurement scenario, provided that the other parties have access to only single copies of the shared quantum state.\\
\textbf{Proposition 1:}
For an $n$-qubit GHZ state subjected to $r$ sequential measurements on a single party (say Alice), characterized by sharpness parameters $\{\lambda_1,\lambda_2,\dots,\lambda_r\}$, the maximal value of the $n$-partite Seevinck--Svetlichny expression \cite{52} obtained by the $r$-th observer is given by
\begin{equation}
S_n^r = 2^{\,n-r}\sqrt{2}\,\lambda_r \prod_{m=1}^{r-1} \left(1 + \sqrt{1 - \lambda_m^2}\right).
\end{equation}
For a general \( n \)-qubit GHZ state, at most two Alice observers can exhibit nonlocality simultaneously while the remaining parties possess only single copies of the shared state under a unilateral sequential sharing scenario( see Appendix~\ref{app:proof1} for details).\\\\
 \begin{table}[h]
 	\caption{Maximal values of $\displaystyle \frac{S_n^{r}}{2^{\,n-1}}$ 
 		for different numbers of sequential Alices. 
 		Values greater than $1$ indicate violation and hence enable 
 		nonlocality sharing.}
 	\label{tab:Sn_over_bound}
 	\begin{ruledtabular}
 		\begin{tabular}{ccc}
 			$r$ & $\max \left( \frac{S_n^{r}}{2^{\,n-1}} \right)$ & Nonlocality Sharing \\ 
 			\hline
 			1 & 1.414 & Yes \\
 			2 & 1.082 & Yes \\
 			3 & 0.944 & No \\
 		\end{tabular}
 	\end{ruledtabular}
 \end{table}\\
 From Table~\ref{tab:Sn_over_bound}, it is evident that violation 
 persists only up to $r=2$, beyond which the quantity 
 $\frac{S_n^{r}}{2^{\,n-1}}$ falls below unity, 
 indicating the absence of sequential nonlocality sharing.
 \begin{tabular}{cccc}
 	$\lambda_i$ &  value of $\lambda_i$ & $\max \left( \frac{S_n^{r}}{2^{\,n-1}} \right)$ & Sharing \\
 	\hline
 	$\lambda_{1}$ & 0.7072 & 1.41 & Yes \\
 	$\lambda_{2}$ & 0.83 & 1.035 & Yes \\
 	$\lambda_{3}$ & 0.9 & 0.932 & No \\
 \end{tabular}\\\\
 \textbf{Nonlocality sharing for mixed state:} We consider the following $n$-qubit mixed state:
\begin{equation}\label{mix}
\rho_{\text{mixed}} = p \vert \varphi \rangle \langle \varphi\vert + (1-p)\frac{I^{\otimes n}}{2^{n}},
\end{equation}
where $ \vert \varphi \rangle = \frac{1}{\sqrt{2}} (\vert 00\cdots 0\rangle + \vert 11 \cdots 1\rangle ) $ and $0 \leq p \leq 1$ denotes the purity parameter characterizing the degree of mixing. For $n = 4$, this mixed state exhibits genuine multipartite nonlocality whenever $ p > \frac{1}{\sqrt{2}}$. Within this regime, both Alice$^1$ and Alice$^2$ can simultaneously demonstrate nonlocal correlations with the remaining parties provided that $p > 0.8848$ and $\lambda_1 > 0.80 $. In contrast, Alice$^1$, Alice$^2$, and Alice$^3$ cannot simultaneously exhibit nonlocality with the other parties. Consequently, under the unbiased sequential measurement formalism, at most two Alices can share nonlocality simultaneously for this mixed state as well.\\
\begin{figure}[htbp]
\centering
\includegraphics[width=0.6\linewidth,trim=1cm 1cm 1cm 1cm, clip]{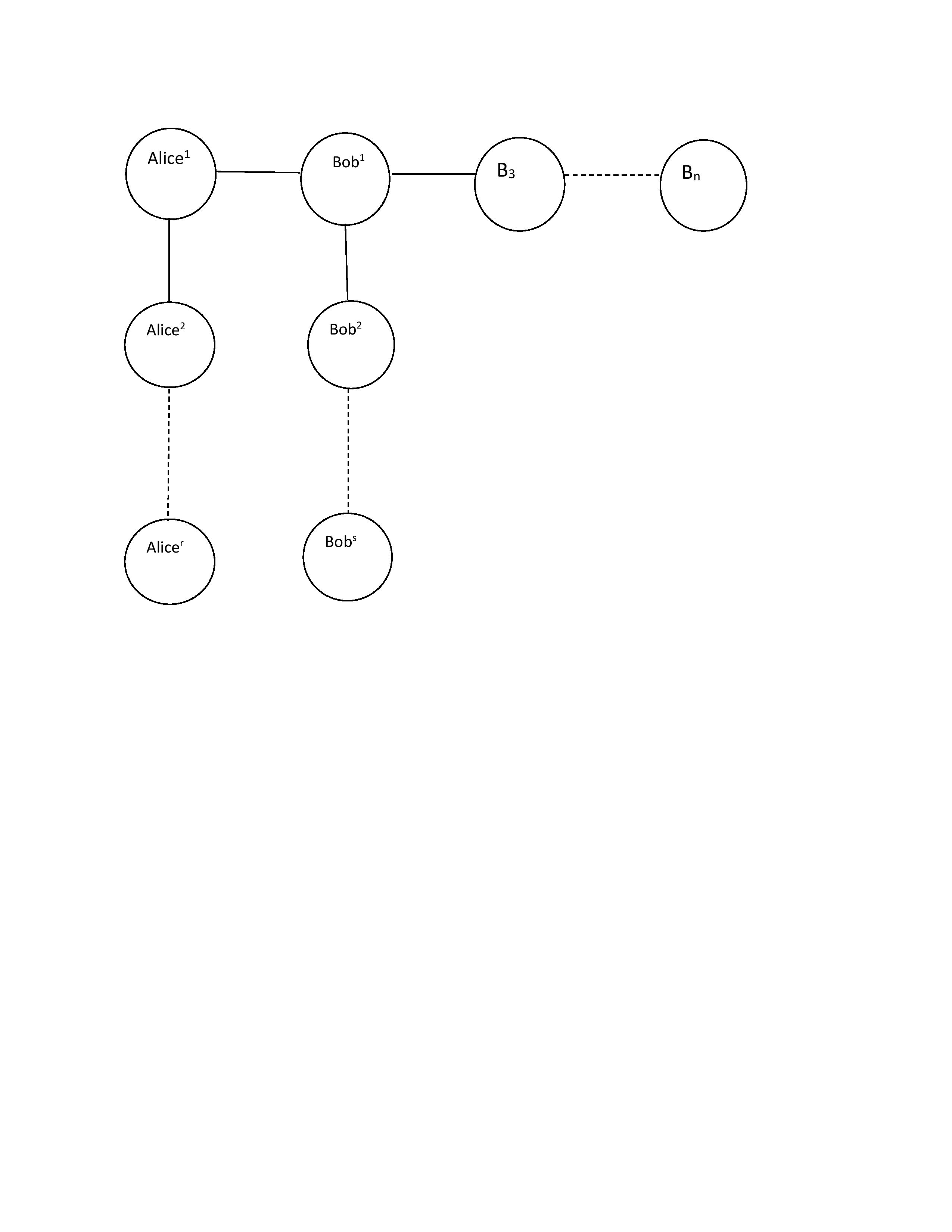}
\captionsetup{skip=-8pt,belowskip=0pt}
\caption{ The figure illustrates the bilateral sharing of multipartite nonlocal correlations, where multiple Alices and multiple Bobs are able to demonstrate nonlocality with the remaining parties. }
\label{sd2}
\end{figure}\\
\subsection{Bilateral Sharing of Multipartite Nonlocality}
The simultaneous sharing of multipartite nonlocality involves exploring the ability of multiple observers to sequentially access and measure nonlocal correlations within a shared quantum system. This study focuses on scenarios where observers, such as sequence of Alice and Bob ( associated with $ B_2 $, say), perform unsharp measurements on their respective subsystems of an \( n \)-qubit GHZ state. The analysis begins with a four-qubit system and extends to general \( n \)-qubit systems. Initially, we consider multiple observers, including Alice and Bob, sharing an \( n \)-qubit GHZ state, with the single copies of the remaining parties. The sequential measurement process involves a sequence of Alice observers, \( \text{Alice}^1 \rightarrow \text{Alice}^2 \rightarrow \text{Alice}^3 \rightarrow \dots \rightarrow \text{Alice}^r \), and a corresponding sequence of Bob observers, \( \text{Bob}^1 \rightarrow \text{Bob}^2 \rightarrow \text{Bob}^3 \rightarrow \dots \rightarrow \text{Bob}^s \), each performing measurements on their respective copies of the state in a sequential manner. In this framework, Alice\( ^1 \) performs a measurement on the first particle and transmits the post-measurement state to Alice\( ^2 \), who, after executing her measurement, forwards the updated state to Alice\( ^3 \), and so forth. Each Alice observer, except for Alice\( ^r \), performs an unsharp measurement on the particle under her possession, introducing minimal disturbance to the quantum system to maximize the possibility of sharing nonlocality with subsequent observers. Similarly, in Bob’s sequence, Bob\( ^1 \) conducts a measurement on his respective particle and passes the post-measurement state to Bob\( ^2 \). Bob\( ^2 \) then measures the particle received from Bob\( ^1 \) and transmits the state to Bob\( ^3 \), continuing this process sequentially. Like Alice's sequence, all Bob observers, except for Bob\( ^s \), execute unsharp measurements.  If any subsequent Alice or Bob ( except Alice$ ^r $ and Bob$ ^s $) performs a sharp measurement on their respective particle, the next observer in their sequence will no longer be able to detect any possible violation of the relevant Bell-type inequality. This is because sharp measurements induce strong collapse, effectively destroying any remaining quantum correlations in the system. In this scenario, Alice and Bob measure different qubits but contribute collectively to the violation of multipartite Bell-type inequalities (e.g., Svetlichny inequalities). The measurement strength of each Alice affects the nonlocal correlations available to each Bob, and vice versa. The degree of unsharpness in the measurements of Alice\( ^m \) and Bob\( ^k \) is characterized by the sharpness parameters \( \lambda_m \) and \( \mu_k \), respectively, where \( m = 1,2, \dots, r \) and \( k = 1,2, \dots, s \). The degree of unsharpness (lower $ \lambda_m, \mu_k $) ensures that more observers can sequentially share nonlocality. \\
The expectation value between Alice$ ^{1} $, Bob$ ^{1} $ and the parties $ B_3, B_4 $ are given by 
\begin{equation}
\overline{EX(i^{1}\overline{i}^{1}j_{3}j_{4})}=\lambda_{1} \mu_{1} \cos(\phi_{i}^{x^{1}}+\phi_{i}^{\overline{x^{1}}}+\phi_{j}^{y_3}+\phi_{j}^{y_4})
\end{equation}
Initially, Alice$ ^{1} $, Bob$ ^{1} $ and the parties $ B_3, B_4 $ generate the following correlation asociated with the Bell-type inequality of Eq. \eqref{dd11} ( utilizing appropriately chosen measurement settings as discussed in section \eqref{sec3}):
\begin{equation}\label{dd15}  
S_4^{1,1} = 8\sqrt{2} \lambda_1 \mu_1,  
\end{equation}  
where the product \( \lambda_1 \mu_1 \) quantifies the combined impact of their respective measurement unsharpness. \\
Hereafter, we consider a scenario where  Alice\(^1\) perform measurement and transmits the post-measurement state to Alice\(^2\), while Bob\(^1\) similarly passes the post-measurement state to  Bob\(^2\). The expectation value between Alice$ ^{2} $, Bob$ ^{2} $ and the parties $ B_3, B_4 $ are given by 
\begin{equation}
\overline{EX(i^{2}\overline{i}^{2}j_{3}j_{4})}=\frac{1}{4}\lambda_{2} \mu_{2} (1+\sqrt{1-(\lambda_{1})^2}+\sqrt{1-(\mu_{1})^2}+\sqrt{1-(\lambda_{1})^2}\sqrt{1-(\mu_{1})^2}) \cos(\phi_{i}^{x^{2}}+\phi_{i}^{\overline{x^{2}}}+\phi_{j}^{y_3}+\phi_{j}^{y_4})
\end{equation}
In this case, the nonlocality parameter for their combined measurements is given by:  
\begin{equation}\label{dd16}  
S_4^{2,2} = 2\sqrt{2} \lambda_2 \mu_2 \left( 1 + \sqrt{1 - (\lambda_1)^2} + \sqrt{1 - (\mu_1)^2} + \sqrt{1 - (\lambda_1)^2} \sqrt{1 - (\mu_1)^2} \right),  
\end{equation}  
where \( \lambda_2 \) and \( \mu_2 \) are the sharpness parameters associated with measurements of Alice\(^2\) and Bob\(^2\), respectively. In order to ensure that Alice$ ^{1} $ and Bob$ ^{1} $ introduce minimal disturbance to the quantum system, thereby optimizing quantum violation for subsequent observers in their respective sequences, the condition \( \lambda_1 \mu_1 > \frac{1}{\sqrt{2}} \) must hold ( from \eqref{dd15}. To determine whether the system still demonstrates nonlocality, we evaluate \( S_4^{2,2} \) under the condition \( \lambda_1 \mu_1 > \frac{1}{\sqrt{2}} \), yielding:  
\[
S_4^{2,2} = 6.96 \lambda_2 \mu_2.  
\]  
In this case Alice$ ^{1} $ and Bob$ ^{1} $ perform unsharp measurements, whereas measurement of Alice$ ^{2} $ and Bob$ ^{2} $ are sharp.
Since this value does not exceed the threshold required to violate the Bell-type inequality, the simultaneous demonstration of four-qubit nonlocality by Alice\(^1\), Alice\(^2\), and Bob\(^1\), Bob\(^2\) is not possible with single copies of the other parties in the system. This result provides a limitations on multilateral sequential sharing of nonlocality. \\
The findings presented above can be extended to quantum systems involving a larger number of qubits. Specifically, in the case of five-qubit and six-qubit systems, the derived expressions for the nonlocality parameters remain valid under analogous conditions. This suggests that the constraints observed in the four-qubit case persist as the system size increases.\\ 
\textbf{Bilateral sequential sharing for five-qubit system:} In the multipartite system comprising multiple Alices, multiple Bobs, and the remaining parties $ B_3, \ldots B_n $, the correlations among Alice$ ^{1} $, Bob$ ^{1} $ and the remaining parties are given by:
\begin{equation} 
S_5^{1,1} = 16\sqrt{2} \lambda_1 \mu_1. 
\end{equation} 
 Alice$ ^{1} $ and Bob$ ^{1} $ can exhibit five-qubit nonlocality simultaneously whenever \( \lambda_1 \mu_1 > \frac{1}{\sqrt{2}} \).  However, we find that  Alice$ ^{1} $, Alice$ ^{2} $, and Bob$ ^{1} $, Bob$ ^{2} $ cannot simultaneously exhibit nonlocality with the remaining parties under this scenario.\\
\textbf{Bilateral sequential sharing for six-qubit system:} In the multipartite system consisting with multiple Alices, multiple Bobs and the remaining parties $ B_3, \ldots B_n $, the correlations among Alice$ ^{1} $, Bob$ ^{1} $ and the remaining parties are given by:
\begin{equation} 
S_6^{1,1} = 32\sqrt{2} \lambda_1 \mu_1. 
\end{equation} 
 Alice$ ^{1} $ and Bob$ ^{1} $ can exhibit six-qubit nonlocality simultaneously for \( \lambda_1 \mu_1 > \frac{1}{\sqrt{2}} \). In this scenario also, Alice$ ^{1} $, Alice$ ^{2} $, and Bob$ ^{1} $, Bob$ ^{2} $ are unable to simultaneously exhibit nonlocality with the remaining parties.\\
 For the mixed state of \eqref{mix}, we find that Alice$ ^{1} $, Alice$ ^{2} $, and Bob$ ^{1} $, Bob$ ^{2} $ are likewise unable to demonstrate nonlocality simultaneously with the remaining parties.\\
\textbf{Proposition 2:}
Consider an $n$-qubit GHZ state shared among $n$ parties, where $r$ sequential observers act on Alice’s qubit with sharpness parameters $\{\lambda_1,\dots,\lambda_r\}$ and $s$ sequential observers act on Bob’s qubit with sharpness parameters $\{\mu_1,\dots,\mu_s\}$. The maximal value of the $n$-partite Seevinck--Svetlichny expression corresponding to the $r$-th Alice and $s$-th Bob is given by:
\begin{equation}
S_n^{r,s} 
= 2^{\,n-r-s+1}\sqrt{2}\,
\lambda_r \mu_s
\left( \prod_{m=1}^{r-1} (1 + \sqrt{1 - \lambda_m^2}) \right)
\left( \prod_{\ell=1}^{s-1} (1 + \sqrt{1 - \mu_\ell^2}) \right).
\end{equation}
We have found that for a general \( n \)-qubit GHZ state, it is not possible for two Alice observers and two Bob observers to simultaneously share nonlocality while the remaining parties each possess a single copy of the system in an unbiased sequential measurement scenario ( for details see Appendix~\ref{app:proof2}).\\
As a result, no further multilateral sequential sharing of nonlocality can be achieved under these conditions. Furthermore, for the special case of a three-qubit system, the phenomenon of bilateral sequential sharing of Svetlichny nonlocality (which characterizes genuine multipartite nonlocality) was previously investigated in \cite{44}, following the formalism introduced in \cite{68}. These studies provide insights into the fundamental limitations of sequential nonlocality sharing in multipartite quantum systems and reinforce the constraints observed in higher-qubit cases, in an unbiased sequential measurement scenario.

\section{ Conclusion}
This study investigates the sharing of genuine multipartite nonlocality for the GHZ states under both unilateral and biltilateral sequential measurement scenarios, employing unbiased unsharp measurements. It is observed that genuine four-qubit nonlocality can be shared by an observer whose multiple copies interact with the state, although the number of such copies is constrained by a specific state-dependent range. For the \( n \)-qubit maximally entangled GHZ state, at most two copies of an observer (e.g., Alice) can share the nonlocality in the unilateral scenario. In scenarios involving multiple observers distributed across different parties, we determine the maximum number of independent observers on $m$ sides who can share genuine multipartite nonlocality with the remaining parties. Notably, in the bilateral scenario, no non-trivial nonlocality sharing is observed, thereby revealing a fundamental limitation of the approach. Consequently, this bilateral result indicates that no additional advantage is attainable in the multilateral scenario. The unbiased unsharp measurement strategy plays a pivotal role in enabling nonlocality sharing in both scenarios. These results emphasize the limitations of equal sharpness measurements and highlight the efficacy of unequal sharpness measurements \cite{30,32,33,69} for recycling qubits to sustain nonlocality. An interesting direction for future research is to examine the sequential sharing of multipartite nonlocality when different parties are characterized by unequal sharpness parameters \cite{30,32}. This work provides valuable insights into the interplay between quantum measurement strategies and the distribution of genuine multipartite nonlocality. It advances the understanding required for optimizing quantum information processing and resource management. In this regard, multipartite nonlocality sharing under projective measurements \cite{62,63,64,65} remains a challenging task.\\
{\bf Acknowledgment:}
S. H. acknowledge B. Paul for his valuable advice.

\appendix

\section{Proof of Proposition 1}
\label{app:proof1}
\begin{proposition}
For an $n$-qubit GHZ state subjected to $r$ sequential measurements on a single party (say Alice), characterized by sharpness parameters $\{\lambda_1,\lambda_2,\dots,\lambda_r\}$, the maximal value of the $n$-partite Seevinck--Svetlichny expression \cite{52} obtained by the $r$-th observer is given by
\begin{equation}
S_n^r = 2^{\,n-r}\sqrt{2}\,\lambda_r \prod_{m=1}^{r-1} \left(1 + \sqrt{1 - \lambda_m^2}\right).
\end{equation}
\end{proposition}

\begin{proof}
We prove the result by induction on the number $r$ of sequential observers.

\textit{Base case ($r=1$):}  
For a single observer performing an unsharp measurement with sharpness parameter $\lambda_1$ on one qubit of the $n$-qubit GHZ state, the correlation tensor scales linearly with $\lambda_1$ due to the Kraus-operator description of the POVM \cite{61,67,70}. The maximal violation of the $n$-partite Seevinck--Svetlichny inequality for the GHZ state is $2^{n-1}\sqrt{2}$ under projective measurements. The effect of unsharpness reduces this by a factor $\lambda_1$, yielding
\[
S_n^1 = 2^{n-1}\sqrt{2}\,\lambda_1,
\]
which agrees with the claimed formula.

\textit{Inductive step:}  
Assume that after $(r-1)$ sequential observers, the correlation function relevant for the $(r-1)$-th observer is given by
\[
S_n^{r-1} = 2^{\,n-(r-1)}\sqrt{2}\,\lambda_{r-1} \prod_{m=1}^{r-2} \left(1 + \sqrt{1 - \lambda_m^2}\right).
\]

Now consider the action of the $r$-th observer, who performs an unsharp measurement with sharpness $\lambda_r$ on the same qubit. The post-measurement state (averaged over outcomes) is obtained via the Lüders update rule \cite{61,67}:
\[
\rho \;\mapsto\; \sum_{i=\pm} K_i \rho K_i^\dagger,
\]
where $K_{\pm} = \sqrt{E_{\pm}^{\lambda_{r}}}$ and
\[
E_{\pm}^{\lambda_{r}} = \frac{1}{2}\left(\mathbb{I} \pm \lambda_r \overrightarrow{n}.\overrightarrow{\sigma}\right).
\]

A direct evaluation shows that, at the level of the correlation tensor, each prior correlation component is transformed multiplicatively by the factor
\[
\frac{1 + \sqrt{1 - \lambda_{r-1}^2}}{2}.
\]
Physically, this reflects the fact that the unsharp measurement partially preserves coherence in the GHZ subspace while attenuating correlations.

Thus, the cumulative effect of $(r-1)$ sequential measurements is to introduce the factor
\[
\prod_{m=1}^{r-1} \frac{1 + \sqrt{1 - \lambda_m^2}}{2}.
\]

Finally, the $r$-th observer contributes an additional factor $\lambda_r$ (as in the base case), corresponding to the linear scaling of correlations under the final unsharp measurement.

Combining all contributions, and noting that the Seevinck-Svetlichny expression contains an overall normalization factor $2^{n-1}$, we obtain:
\[
S_n^r = 2^{n-1}\sqrt{2} \,\lambda_r 
\prod_{m=1}^{r-1} \frac{1 + \sqrt{1 - \lambda_m^2}}{2}.
\]

Rewriting,
\[
S_n^r = 2^{\,n-r}\sqrt{2}\,\lambda_r \prod_{m=1}^{r-1} \left(1 + \sqrt{1 - \lambda_m^2}\right),
\]
which completes the induction.
\end{proof}
It is observed that Alice$^{1}$ and Alice$^{2}$ can simultaneously demonstrate $n$-partite nonlocality provided $\lambda_{1} > \frac{1}{\sqrt{2}}$ and $\lambda_{2} > \frac{\sqrt{2}}{1 + \sqrt{1 - \lambda_{1}^{2}}}$. However, the simultaneous demonstration of genuine $n$-partite nonlocality by Alice$^{1}$, Alice$^{2}$, and Alice$^{3}$ is not possible. It is worth noting that, in each sequential scenario, the final Alice always performs sharp measurements.

\section{Proof of Proposition 2}
\label{app:proof2}
\begin{proposition}
Consider an $n$-qubit GHZ state shared among $n$ parties, where $r$ sequential observers act on Alice’s qubit with sharpness parameters $\{\lambda_1,\dots,\lambda_r\}$ and $s$ sequential observers act on Bob’s qubit with sharpness parameters $\{\mu_1,\dots,\mu_s\}$. The maximal value of the $n$-partite Seevinck--Svetlichny expression corresponding to the $r$-th Alice and $s$-th Bob is given by:
\begin{equation}
S_n^{r,s} 
= 2^{\,n-r-s+1}\sqrt{2}\,
\lambda_r \mu_s
\left( \prod_{m=1}^{r-1} (1 + \sqrt{1 - \lambda_m^2}) \right)
\left( \prod_{\ell=1}^{s-1} (1 + \sqrt{1 - \mu_\ell^2}) \right).
\end{equation}
\end{proposition}
\begin{proof}
we explicitly note that the sequential measurement processes on Alice’s and Bob’s qubits act on distinct subsystems, and hence their corresponding quantum channels commute and factorize at the level of the correlation tensor. As a result, the overall transformation of the correlation tensor is given by the product of the independent contributions from the two chains.\\
Therefore, \textbf{Proposition 2} follows directly by applying \textbf{Proposition 1}\eqref{app:proof1} separately to Alice’s and Bob’s sequential measurement sequences and combining the resulting factors. This yields the desired expression in a straightforward manner.
\end{proof}
It is observed that Alice$^{1}$ and Bob$^{1}$ can simultaneously demonstrate $n$-partite nonlocality when $\lambda_{1}\mu_{1} > \frac{1}{\sqrt{2}}$. In contrast, the simultaneous demonstration of $n$-partite nonlocality by Alice$^{1}$, Alice$^{2}$ and Bob$^{1}$, Bob$^{2}$ is not possible.\\


\end{document}